\documentclass[12pt]{article}

\usepackage{epsfig}
\usepackage{cite}
\usepackage{amsmath, amssymb, amsfonts}
\usepackage{color}
\usepackage{latexsym}  
\usepackage{graphicx}
\usepackage{cancel}
\usepackage[colorlinks,bookmarks]{hyperref}
\hypersetup{pdfpagemode=UseNone, pdfstartview=FitH, linkcolor=blue, citecolor=red, urlcolor=blue}

\def\be{\begin{equation}}
\def\ee{\end{equation}}
\def\ba{\begin{eqnarray}}
\def\ea{\end{eqnarray}}

\def\bdm{\begin{displaymath}}
\def\edm{\end{displaymath}}

\def\bq{\begin{quote}}
\def\eq{\end{quote}}

 at 10truept

\newcommand{\bea}{\begin{eqnarray}}
\newcommand{\eea}{\end{eqnarray}}

\newcommand{\bi}{\begin{itemize}}
\newcommand{\ei}{\end{itemize}}

\newcommand{\beq}{\begin{equation}}
\newcommand{\eeq}{\end{equation}}
\newcommand{\beqa}{\begin{eqnarray}}
\newcommand{\eeqa}{\end{eqnarray}}

\def\12{{1 \over 2}}

\def\ltap{\ \raise.3ex\hbox{$<$\kern-.75em\lower1ex\hbox{$\sim$}}\ }
\def\gtap{\ \raise.3ex\hbox{$>$\kern-.75em\lower1ex\hbox{$\sim$}}\ }
\def\gl{\ \raise.5ex\hbox{$>$}\kern-.8em\lower.5ex\hbox{$<$}\ }
\def\roughly#1{\raise.3ex\hbox{$#1$\kern-.75em\lower1ex\hbox{$\sim$}}}

\begin{document}

\thispagestyle{empty}
\begin{flushright}
June 2026
\end{flushright}
\vspace*{1.1cm}
\begin{center}

{\Large \bf Localization of Chiral Electromagnetic Waves on}
\vskip.3cm
{\Large \bf Thick Axion Domain Walls} 

\vspace*{1.35cm} {\large
Nemanja Kaloper\footnote{\tt
kaloper@physics.ucdavis.edu} }\\
\vspace{.5cm}
{\em QMAP, Department of Physics and Astronomy, University of
California}\\
\vspace{.05cm}
{\em Davis, CA 95616, USA}\\

\vspace{1.25cm} ABSTRACT
\end{center}
We analyze Maxwell theory coupled to an axion domain wall as 
a spectral boundary value problem. We find that a finite-width axion domain wall generically 
supports a localized normalizable chiral electromagnetic mode with linear, gapless 
dispersion. This mode arises from helicity-dependent coupling sourced by the axion gradient: 
one polarization experiences an effective attractive potential and forms a bound state, 
while the opposite polarization is repelled. The existence of this chiral surface photon is robust 
over a wide regime of wall structures and axion masses. Our result shows that axion domain 
walls generically support a localized chiral photon that has been missed in previous analyses.

\vfill \setcounter{page}{0} \setcounter{footnote}{0}

\vspace{1cm}

\newpage

\section{Introduction}

Axion electrodynamics is a simple and widely studied framework in which 
topological couplings modify the dynamics of electromagnetic fields propagating in
nontrivial backgrounds \cite{Wilczek:1987mv,Qi:2008ew,Essin:2008rq,Karch:2009sy}. 
In particular, spatial variations of an axion field 
induce helicity-dependent couplings that act directly on photon polarization. 
These effects are most commonly discussed in the context of birefringence, 
where electromagnetic waves propagating through an axion 
background undergo polarization rotation 
\cite{Huang:1985tt,Harari:1992ea,Komatsu:2022nvu,Raffelt:1987im}. 
In this work we show that the exact same 
mathematical structure which describes birefringence can be reinterpreted as a spectral 
boundary value problem for a planar domain wall. An incident bulk
electromagnetic field scattering on the wall then becomes a diagnostic to search for 
a propagating electromagnetic mode localized 
on the wall. In this sense, electromagnetic mode localization 
and birefringence are but two aspects of a single underlying mechanism 
\cite{Kaloper:2026slg,Kaloper:2026ygk,Kaloper:2026auc}.

To make this connection precise, we review the ultrathin 
wall limit \cite{Kaloper:2026auc}, in 
which the axion profile reduces to a Chern--Simons interface, which supports 
a $\delta$-function Chern--Simons interaction in the $3+1$ 
bulk \cite{Deser:1981wh,Zanelli:2010zz}. In this regime the 
electromagnetic problem may be solved using a Lippmann--Schwinger 
formulation. We can identify the localized mode as a pole in the 
scattering amplitude of bulk photons, even without explicitly constructing 
those modes. The singular properties of the scattering amplitude provide a direct,
physically transparent diagnostic of localization, independent of the 
details of bound-state wavefunctions. The underlying mechanism 
is chiral, precisely confirming the results found by 
direct construction in the ultrathin limit \cite{Kaloper:2026auc}. 
The wall couples with opposite sign to the two photon helicities. This yields an effective 
attractive regime for one polarization and a repulsive regime for the other. 
As a result, only a single helicity supports a normalizable surface excitation.

We then extend this analysis to axion domain walls of finite thickness, described 
by smooth background profiles \cite{Huang:1985tt}. We focus on the concrete example when the
axion profile interpolates between different vacua of the standard harmonic
axion potential $\propto \cos(\phi/f_\phi)$. In this case the electromagnetic field propagates 
in a spatially varying chiral background induced by $\vec \nabla \phi$, 
reducing the problem to a one-dimensional system with a helicity-dependent 
integrable potential \cite{Kaloper:2026ygk}. The eigenvalue problem is 
continuously connected to the thin-wall limit, ensuring 
that the localized mode persists for arbitrary wall thickness. 

Although we do not 
construct the bound-state solutions exactly, we establish the existence 
of a single chiral surface photon with linear, gapless dispersion. To show this we use the analyticity of the 
scattering matrix of bulk modes and its smooth interpolation of the $\delta$-function, 
at frequencies below the axion mass, $\omega < m_\phi$. The axion mass
$m_\phi$ naturally serves as an effective UV cutoff
for the localized mode perturbation theory. In this regime, the localized mode should be 
understood as a low-energy bound-state excitation of the effective theory, 
which is continuously connected to the exact thin-wall solution. 
The localized electromagnetic mode dwells 
precisely in the regime where the domain wall admits a consistent effective 
description as a coherent background. For those modes, the wall acts as a
well-defined ``obstacle" to electromagnetic propagation. 
Our analysis demonstrates 
that in this regime electromagnetic localization on axion domain walls is a robust and axion
profile-independent consequence of the chiral topological coupling. It would be interesting to 
search for the spectrum of these modes in explicit form.

\section{Localized Electromagnetic Waves on Ultrathin Walls}

In this section we summarize the key results of \cite{Kaloper:2026auc}, 
where we showed that Maxwell theory in the presence of a codimension-$1$ 
Chern--Simons interface has a propagating electromagnetic mode 
localized on the interface. We focus on the physical structure of the solution, 
referring the reader to  \cite{Kaloper:2026auc} for the details of derivation. 
We begin with Maxwell theory coupled to a pseudoscalar background field 
$\theta(x)$ through the Chern--Simons density,
\be
S = -\frac{1}{4}\int d^4x\, F_{\mu\nu}F^{\mu\nu}
- \frac{1}{4}\int d^4x\, \theta(x)\,
\epsilon^{\mu\nu\lambda\sigma} F_{\mu\nu}F_{\lambda\sigma} \, .
\label{actioncs}
\ee
When $\theta$ is constant, the second term is a total derivative and does not affect 
local photon propagation. Nontrivial dynamics arise only from regions where 
$\theta$ varies. The simplest example is when $\theta$ is piecewise constant,
\be
\theta(z) = \theta_- + \Delta\theta\,\Theta(z) \, ,
\label{thesum}
\ee
so that $\partial_z \theta = \Delta\theta\,\delta(z)$ localizes the interaction 
to a planar interface at $z=0$. This configuration describes an infinitesimally
thin interface supporting electromagnetic Chern--Simons term, separating the
empty space vacua with constant $\theta_\pm$.

Varying the action and ignoring any charges
and currents, electric or magnetic,  yields the modified Maxwell equations
\be
\partial_\mu \Bigl(
F^{\mu\nu}
+ \theta(x)\,\epsilon^{\mu\nu\lambda\sigma} F_{\lambda\sigma}
\Bigr) = 0 \, ,
\label{covsum}
\ee
which reduce to ordinary vacuum electrodynamics 
away from the interface. Using 
the electric and magnetic fields,
\be
~~~~~~ E^i = F^{0i}\, , \qquad \qquad B^i = \frac{1}{2}\epsilon^{ijk}F_{jk} \, ,
\ee
the field equations take the form
\ba
&&~~~~~~~\vec \nabla\cdot\vec E = -\Delta\theta\,\delta(z)\,B_z \, , \qquad \qquad 
\vec \nabla\times\vec B - \partial_t\vec E 
= \Delta\theta\,\delta(z)\,\hat z\times\vec E \, , \nonumber\\
&&~~~~~~~\vec \nabla\cdot\vec B = 0 \, , \qquad \qquad  \qquad \qquad  ~~\,
\vec \nabla\times\vec E + \partial_t\vec B = 0 \, .
\label{maxwellsum}
\ea
The equations in the first line of \eqref{maxwellsum} imply 
nontrivial matching conditions across the interface, which 
encode the physical effect of the Chern--Simons coupling. Since the sources are
proportional to $\delta(z)$, they are naturally interpreted as boundary conditions
for $\vec E$ and $\vec B$ on the interface, 
just like in the case of Fresnel refraction \cite{Kaloper:2026slg}. 
 
We look for solutions corresponding to waves propagating along the interface 
and localized in the direction normal to the interface.  Hence along the interface we
use the ans\"atz
\be
\vec E = \vec{\cal E}(z)\,e^{i(\vec k_\parallel\cdot\vec x_\parallel-\omega t)} \, , 
\qquad \qquad 
\vec B = \vec{\cal B}(z)\,e^{i(\vec k_\parallel\cdot\vec x_\parallel-\omega t)} \, .
\label{ansatzint}
\ee
Away from the interface, the fields satisfy the same vacuum wave equation on both sides
of the interface. Substituting \eqref{ansatzint} into the bulk Eqs. \eqref{maxwellsum} and combining
the results yields 
\be
\bigl(\partial_z^2 + \omega^2 - k_\parallel^2\bigr)\vec{\cal E}_\pm(z) = 0 \, ,
\qquad \qquad
\bigl(\partial_z^2 + \omega^2 - k_\parallel^2\bigr)\vec{\cal B}_\pm(z) = 0 \, ,
\label{wavebulk}
\ee
where $k_\parallel^2 = \vec k_\parallel^{\,2}$, 
and obeys the same bulk dispersion relation
away from the interface, because the bulk on 
either side satisfies standard vacuum Maxwell equations. 
The electromagnetic fields on different sides of the 
interface are not the same, since the boundary conditions come 
from the Chern--Simons term, that is not invariant 
under parity reflection across the interface. 
Localization to the wall is then ensured by the exponential decay,
\be
\vec{\cal E}_\pm(z),\;\vec{\cal B}_\pm(z) \propto e^{-\kappa |z|} \, ,
\qquad \qquad
\kappa^2 = k_\parallel^2 - \omega^2 \, . 
\label{normansatz}
\ee
The full analysis of enforcing boundary conditions on the eigenmodes 
and determining their exact form is given in necessary detail in \cite{Kaloper:2026auc}.
Here we merely quote the final result and review its properties. 
For definiteness, we take $\Delta\theta>0$, which implies that the positive-frequency 
localized mode is 
left-handed. The resulting electromagnetic fields 
which include the field components everywhere,
along the interface and normal to it, are
\ba
\vec{\cal E}_\pm(z) &=& {\cal E}_0
\left[ \vec e_{\mathrm{L}} \;\pm\;
\frac{i}{\kappa} (\vec k_\parallel\cdot\vec e_{\mathrm{L}})\,\hat z \right]
e^{-\kappa|z|} \, , \nonumber \\
\vec{\cal B}_\pm(z) &=&
\frac{{\cal E}_0}{\omega}
\left[
\vec k_\parallel\times\vec e_{\mathrm{L}}
\;\pm\; \kappa\,\vec e_{\mathrm{L}}
\right] e^{-\kappa|z|} \, ,
\label{allfields}
\ea
where the chiral eigenbasis vectors are 
$\vec e_{\mathrm{L,R}} = (\hat x \pm i\hat y)/\sqrt{2}$ and
the wave vector $\vec k_\parallel$ is the conserved wave momentum. 
These solutions describe a wave localized on the interface and propagating parallel 
to it.

Consistency of the boundary conditions fixes the relation between $\kappa$ 
and $\omega$,
\be
\kappa = \frac{\Delta\theta\,\omega}{2} \, ,
\ee
which, when combined with the bulk dispersion relation given in Eq. \eqref{normansatz}, yields
\be
\omega^2 =
\frac{k_\parallel^2}{1 + \frac{(\Delta\theta)^2}{4}} \, .
\label{disp_summary}
\ee
Despite the wall-localized Chern-Simons terms, this dispersion relation is linear 
and gapless, with both phase and group 
velocities given by
\be
v_{\rm ph} = v_{\rm g} =
\frac{1}{\sqrt{1+\frac{1}{4}(\Delta\theta)^2}} < 1 \, .
\ee
The resulting mode is a massless single-helicity 
surface photon exponentially localized on the 
Chern--Simons interface. Its existence can be understood as a consequence of 
the chiral nature of the interaction: the interface acts as an attractive 
$\delta$-function potential for one helicity and repulsive for the other. The 
Chern--Simons term is a chiral derivative interaction, instead of a mass. 
This produces a unique normalizable solution bound to the interface.

Several features are noteworthy. The mode is gapless and nondispersive, with 
a frequency-independent propagation speed set only by $\Delta\theta$. It does 
not rely on any ambient medium or geometric confinement, and persists in 
otherwise empty vacuum. The energy flux is localized near the interface and 
flows entirely tangentially, confirming that the excitation is genuinely 
bound and does not radiate into the bulk \cite{Kaloper:2026auc}.

It was also noted in \cite{Kaloper:2026auc} that this mode can also be identified as a pole in the 
scattering amplitude of bulk photons, providing a complementary spectral 
diagnostic of its existence. We now elaborate this point, since we will use its extension
to finite width wall cases to show that a localized chiral electromagnetic mode hides
in those walls too.

\section{Lippmann-Schwinger Bound State Detection}

We can identify the presence of a localized chiral electromagnetic wave on the 
Chern-Simons interface independently, using the Lippmann--Schwinger formulation
for scattering of bulk modes. Rather than solving directly for bound-state 
wavefunctions, the Lippmann--Schwinger approach 
treats the problem as a scattering problem and detects localized modes as 
spectral singularities of the scattering matrix elements. This provides a complementary 
and conceptually distinct diagnostic of the same physical excitation 
\cite{Kaloper:2026ygk}. The scattering formalism has been developed for the 
application to CMB birefringence problem recently 
(see also \cite{Ganoulis:1986rd,Favitta:2023hlx,Blasi:2024xvj}), so we
can transfer much of the technical framework here. 

The propagation of bulk electromagnetic modes through
the wall can be set up as the scattering of a normally-incident electromagnetic 
wave coming from bulk infinity, moving in the $z$-direction, encountering the
wall, and being partially scattered and partially transmitted. This picture is valid 
when the wall curvature radius is much larger than the wavelength of the 
incident wave. The final step in setting it up involves a boost in the $z-t$ plane which picks the
incident wave frame so that the incident angle is at 90 degrees \cite{Huang:1985tt,Harari:1992ea}. 
Since in this case $\vec k \parallel \hat z$, we have $\vec k \cdot \vec x = k z$, and the
wave phase only changes in the $z$-direction. 

In this case the problem can be formulated as the evolution of 
the electromagnetic $4$-vector potential, where we can consistently pick Lorentz gauge and
then further restrict it to the axial transverse gauge $A^\mu = (0, A^x, A^y, 0)$ off 
and on the wall 
\cite{Huang:1985tt,Ganoulis:1986rd,Favitta:2023hlx,Blasi:2024xvj,Kaloper:2026ygk}. 
The relevant field equation for the transverse vector 
field components, which follows from \eqref{covsum}, is 
\be
(\partial_z^2 + \omega^2)\,\vec A_\perp(z)
= \Delta\theta\,\omega\,\delta(z)\,\sigma_2\,\vec A_\perp(z) \, .
\label{vecpot}
\ee
Here $\sigma_2$ is a Pauli spin matrix. 
Since $\vec E = - \partial_t \vec A$ and $\vec B = \vec \nabla \times \vec A$, the
field equations for incident $\vec E$ and $\vec B$ are the same. 
For later convenience, we will stick with 
$\vec A_\perp(z)$ variables.  

The interface appears here as a localized matrix-valued potential acting on the 
two transverse components. We can diagonalize 
this system by going to circular polarizations
\be A_\pm(z)
= \frac{1}{\sqrt{2}}\Bigl(A_\perp^x(z) \pm i A_\perp^y(z)\Bigr) \, .
\label{circcomp}
\ee
For these variables, the differential equation \eqref{vecpot} translates to
\be
(\partial_z^2 + \omega^2)\, A_\pm
= \pm \Delta\theta\,\omega\,\delta(z)\, A_\pm \, .
\label{vecpotchi}
\ee
We can invert Eq. \eqref{vecpotchi} and write the integral Lippmann-Schwinger equation, by inverting 
$(\partial_z^2 + \omega^2)$ using the outbound Green's function
$G(z - z') = -\frac{i}{2\omega}\, e^{\,i\omega |z - z'|}$. This gives\footnote{Our sign conventions here 
for Fourier transform follow the standard prescription where evolution forward in time 
proceeds from left to right, in contrast to the convention which we deployed in 
\cite{Kaloper:2026ygk} where evolution proceeded in the opposite direction, that fitted
better the cosmological application. The transition from one to the other is straightforward, 
and may involve adjustment of the integration contour choice.}
\be
A_\pm(z) = e^{i\omega z} 
- \frac{i}{2\omega} \int_{-\infty}^{\infty} dz'\,
e^{\,i\omega |z - z'|}\, V_\pm(z')\, A_\pm(z') \, ,
\label{LSdelta}
\ee
with the potential $V_\pm(z') = \pm \Delta\theta\,\omega\,\delta(z')$. Since the potential
is a $\delta$-function, the integral can be done easily. At this point we should flag 
a very important subtlety concerning analytical
continuation of the operator eigenvalues $\omega^2 = \lambda$ due to the factor of $\omega$ 
which normalizes $\delta$ in $V_\pm$. The former is the eigenvalue 
of the operator $\partial_z^2$, while the latter comes from the 
time derivative of the gauge field
which yields the right hand side term in \eqref{vecpot}. Thus when inverting \eqref{vecpotchi} and
analytically continuing eigenvalues of $\partial_z^2$, we keep the explicit factor of $\omega$
in the numerator fixed. 

Now for the $\delta$-potential the integral Lippmann-Schwinger equation \eqref{LSdelta} 
collapses to 
\be
A_\pm(z) = e^{i\sqrt{\lambda}  z} \;\mp\; 
\frac{i\Delta\theta \omega}{2\sqrt{\lambda} }\, e^{i\sqrt{\lambda} |z|}\, A_\pm(0)\,.
\label{Aexplicit}
\ee
and so after a brief calculation at $z=0$, by matching terms we find 
\be
A_\pm(0)=\frac{1}{1 \pm \frac{i\Delta\theta \omega}{2\sqrt{\lambda}}} \, .
\label{A0sol}
\ee
Combining this and the solution \eqref{Aexplicit}, we 
find the transmission and reflection coefficients,
\be
T_\pm = \frac{1}{1 \pm \frac{i\Delta\theta \omega}{2\sqrt{\lambda}}}\, , \qquad  \qquad
R_\pm = \mp i \frac{\Delta\theta \omega}{2\sqrt{\lambda}} 
\frac{1}{1 \pm \frac{i\Delta\theta \omega}{2\sqrt{\lambda}}} \, . 
\label{transrefl}
\ee

We immediately see that the scattering matrix elements in \eqref{transrefl} have a pole 
at $1 \pm \frac{i\Delta\theta \omega}{2\sqrt{\lambda}}$. This corresponds to the 
eigenvalue $\sqrt{\lambda} = i \kappa$ where
\be
\kappa = \mp \frac{\Delta\theta \omega}{2} \, . 
\label{kappapole}
\ee
For $\Delta \theta > 0$ and $\omega > 0$, the solution $\kappa < 0$ corresponds to the 
right-handed helicity which is delocalized; however the other solution is the localized 
left-handed helicity found in \cite{Kaloper:2026slg}; 
indeed, when $\kappa>0$ the ``scattered" mode
in \eqref{Aexplicit} is $\propto e^{-\kappa |z|}$. 

This extraction of the localized solution from the 
scattering matrix elements follows from the general theory 
of the analytic structure of operator inversion. This is the backbone of our derivation and so 
we link it to the general results. First we analytically continue the eigenvalues $\omega^2$ 
to complex plane, by extending $\omega^2 \rightarrow \lambda$,
and Fourier-transform variables like \eqref{vecpotchi} to momentum domain for $z$ by
setting $A_\pm = \int \frac{dq}{2\pi} \, {\cal A}_\pm(q) \, e^{iqz}$. 
Then after formally inverting $(\partial_z^2 + \lambda)^{-1}$ in Eq. \eqref{vecpotchi} and
dropping the free wave, 
\be
{A}_\pm(z) 
=  \pm \Delta\theta\,\omega\, A_\pm(0) \, \int_{-\infty}^{\infty} \, 
\frac{dq}{2\pi} \, \frac{e^{i q z}}{\lambda - q^2}  \, . 
\label{foutra}
\ee
Now we look for the poles of the Fourier kernel 
$1/(\lambda - q^2)$ in the complex $q$-plane, whose 
physical interpretation comes from the integration contour for Fourier transforms, which is 
determined by the location of the poles
and the requirement that the configuration \eqref{foutra}  solves the equations \eqref{vecpot}
with correct boundary conditions. 

If $\lambda>0$, the poles of the integrand are at $q = \pm \sqrt{\lambda}$, and  
the Fourier integral produces oscillatory modes $e^{\pm i \sqrt{\lambda}\, z}$, 
which describe propagating waves in the bulk. We must pick 
the contours which run along the real $q$ axis, and close it by deforming it around the poles 
and at infinity in order to implement the choice of incoming and outgoing 
boundary conditions, as is usual in 
scattering theory \cite{Newton:1982qc,messiah}.

However, $1/(\lambda - q^2)$ also has poles 
along the imaginary $q$ axis which occur for $\lambda < 0$.
These are precisely the poles of most interest to us, since some represent the 
localized modes we are after. Indeed picking $\lambda = -\kappa^2$ with $\kappa > 0$,
the poles $q = \pm i\kappa$ dictate that to evaluate \eqref{foutra} now we need to close the contour 
in either the upper or lower half-plane, depending on the sign of $z$. For $z>0$ 
we close the contour in the upper half-plane, enclosing the pole at $q = +i\kappa$, 
while for $z<0$ we close in the lower half-plane, enclosing the pole at $q = -i\kappa$. 
In both cases, the residue theorem yields the exponentially decaying result for both sides,
\be
A_\pm(z) \propto e^{-\kappa |z|} \, .
\label{impole}
\ee
The other choice of contour closures leads to nonnormalizable solutions.
Hence the analytic continuation of the Fourier kernel into the complex plane is 
precisely the general tool to detect the normalizable solutions at the interface. It is
directly tied to the poles in the scattering amplitude, and the investigation of those
is the shortcut to unveil the spectrum of localized modes a.k.a. bound states.  
Concretely, we see that 
our bound state of light on an ultrathin Chern--Simons layer 
is precisely encoded by a pole of the analytically continued scattering 
amplitude of bulk modes, being located on the 
evanescent sheet of the complex momentum plane. 

Indeed, the kernel $(\lambda - q^2)^{-1}$ 
appearing in momentum space is the Fourier-space 
representation of the resolvent  
\be
R(\lambda) = (\lambda - H_0)^{-1} \, , 
\label{resolvent}
\ee
where $H_0 = -\partial_z^2$ is the free bulk 
operator. The full Lippmann--Schwinger 
equation therefore takes the form
\be
A_\pm = R(\lambda)\, V_\pm\, A_\pm \, ,
\label{oplipsch}
\ee
which in our case involves $V_\pm(z) = \pm \Delta\theta\,\omega\,\delta(z)$. The 
existence of a localized mode is equivalent to the operator 
$1 - R(\lambda)V_\pm$ failing to be invertible. This occurs precisely at values of 
$\lambda$ for which the resolvent develops a pole.

To summarize, the analytic 
structure of the dressed Lippmann-Schwinger propagator 
encodes both the continuum of scattering states and 
any bound states in the theory. The poles of the analytically continued 
resolvent with positive imaginary part correspond to normalizable 
localized eigenfmodes of the full operator 
\cite{Newton:1982qc,messiah}. To find they are present, 
we can consider the scattering amplitude involving only asymptotically free states, 
and explore its analyticity at complex momenta. 
The surface electromagnetic mode we constructed
in \cite{Kaloper:2026slg} is not an independent add-on to the 
spectrum, but is already encoded in the scattering problem itself. The 
Lippmann--Schwinger approach merely provides a 
diagnostic to reveal it, and its avatars in other systems. 
We can therefore apply this method to other settings 
including finite width axion domain walls.

\section{Low-Energy CP-Odd Theory and its Domain Walls}

Turning to the finite width axion domain walls, 
we first establish a relationship between them and the 
ultrathin domain walls from the previous sections. 
Focusing on the CP-odd sector relevant for photon propagation,
which is Lorentz- and gauge-invariant, we can define the effective field theory in flat 
spacetime \cite{Kaloper:2025goq,Kaloper:2025wgn,Kaloper:2025upu}, 
that schematically is given by the action 
\be
S \;\supset\; \int d^4x \left[
-\frac{1}{4}F_{\mu\nu}F^{\mu\nu}
+ \frac{1}{2}(\partial \phi)^2
- V_{\tt I}\!\left(\frac{\phi}{f_\phi} + \frac{\hat\theta}{2\pi}\right)
- \frac{1}{2}\Theta^2
- \frac{\zeta}{4!{\cal M}^2}\,\Theta\,
\epsilon^{\mu\nu\lambda\sigma}F_{\mu\nu}F_{\lambda\sigma}
\right] \, ,
\label{action}
\ee
where we have defined the CP-odd combination
\be
\Theta = \sqrt{\cal X}\frac{\phi}{f_\phi} + {\cal H} \, ,
\label{bigth}
\ee
and suppressed the sector of the action describing the discharge channel for ${\cal H}$. 
Here $\phi$ is an axion-like field, ${\cal H}$ is the magnetic dual of the 
$4$-form field strength which monodromizes the 
axion, and $\zeta$ controls the strength of the induced 
Chern--Simons coupling to electromagnetism. The scale ${\cal M}$ sets the 
ultraviolet cutoff of the effective field theory description.

At energies below ${\cal M}$ the heavy degrees of freedom may be integrated out, 
and the interaction reduces to an effective axion electrodynamics, where
electromagnetism couples to a pseudoscalar background through
\be
{\cal L}_{\rm int} \;\sim\; \theta(x)\, \epsilon^{\mu\nu\lambda\sigma}F_{\mu\nu}F_{\lambda\sigma} \, ,
\label{CSth}
\ee
with $\theta(x)$ identified with the low-energy CP-odd field configuration in \eqref{bigth}. 
As in the sharp-interface limit discussed above, only spatial gradients of 
$\theta$ have physical effects on electromagnetic propagation. The difference is that the axion 
$\phi$ may be a dynamical field with a mass below the cutoff ${\cal M}$ which means
that it can vary smoothly. 

The vacuum structure of the theory is discrete, reflecting underlying low-energy 
shift symmetries of the axion and quantization of the dual four-form. In particular, 
the axion has a discrete shift symmetry
\be
\frac{\phi}{f_\phi} \;\to\; \frac{\phi}{f_\phi} + n \, , \qquad n \in \mathbb{Z} \, .
\label{vacuadist}
\ee
Concurrently ${\cal H}$ shifts in discrete units set by membrane charge. These 
symmetries identify a lattice of degenerate vacua, differing by discrete values 
of the CP-odd field $\theta$. The flux ${\cal H}$ may be effectively frozen at scales 
below the cutoff ${\cal M}$ depending on the charges and tensions of the membranes 
which can discharge ${\cal H}$. Simply put, a 
discharge of ${\cal H}$ by membrane nucleations
could be too slow to matter. Conversely we may also consider a limit where the
axion $\phi$ is fixed for all practical intents and purposes: loosely, this corresponds to
the limit when the axion mass is above the cutoff ${\cal M}$, and its tunneling rates
to move from one vacuum to another are too slow. In this limit, the dominant vacuum transitions
could be mediated by ${\cal H}$ discharges.

In any case, if we are agnostic about the specifics of dominant discharges, we can simply
focus on the generic case where transitions between distinct vacua are mediated 
by domain walls across which 
$\theta(x)$ interpolates between different constant values. This means
that in general we allow thin-wall limit discussed previously, but also 
allow for smooth configurations,
\be
\theta(z) = \theta_0 + \Delta\theta\, f(z) \, ,
\ee
where $f(z)$ is a monotonic profile that varies over a finite width $L$ and 
approaches asymptotic constant vacuum values as $z \to \pm\infty$.

The finite-width domain walls are therefore a continuous interpolation between 
distinct topological sectors of the theory. As we will see they preserve the 
essential chiral structure of the electromagnetic coupling. As a result they admit 
a natural generalization of the surface-localized electromagnetic modes which 
we identified in the $\delta$-function limit, both by explicit construction of these
modes and by the analysis of the scattering amplitude of bulk
modes. In what follows, to keep the analysis as simple as possible,
we will refrain from solving explicitly for bound states in the 
smooth axion wall background. Instead, as elaborated above, we will prove their existence and 
investigate some of their properties through the 
spectral structure of photon propagation in the wall profile.

\section{Field Equations and Axion Domain Wall Background}

Let us freeze the $4$-form sector and redefine the axion to absorb 
the ${\cal H}$ flux, $\varphi = \phi + f_\phi\,\frac{{\cal H}}{{\cal M}^2}$ and 
$\tilde\theta = \hat\theta - 2\pi\,\frac{{\cal H}}{{\cal M}^2}$. The low energy 
axion-Maxwell effective action becomes 
\be
S = \int d^4 x \Bigl\{
-\frac{1}{4} F_{\mu\nu} F^{\mu\nu}
- \frac{\zeta}{4! f_\phi}\,
\varphi\,\epsilon^{\mu\nu\lambda\sigma} F_{\mu\nu}F_{\lambda\sigma}
+ \frac12 (\partial\varphi)^2
- V_{\tt eff}(\varphi)
\Bigr\} \, .
\label{effaction}
\ee
where 
$V_{\tt eff}(\varphi) =
V_{\tt I}\!\left(\frac{\varphi}{f_\phi} + \frac{\tilde\theta}{2\pi}\right)
+ \frac{{\cal M}^2}{2 f_\phi^2}\,\varphi^2$ is the effective axion 
potential which includes both the instanton contributions
and the top-form mass term correction. To focus on the 
standard axion domain wall description, we 
further take the limit where ${\cal M} \ll \mu^2$, so that the 
$4$-form contributions are completely ignorable. This leaves us
with the ``canonical" axion theory in the dilute instanton gas approximation, 
\be
V_{\tt eff} \rightarrow V_{\tt I} = \mu^4\!
\left[1-\cos\!\left(2\pi\frac{\varphi}{f_\phi} + \tilde\theta\right)\right] \, .
\label{sineG}
\ee
While we will work with this axion wall profile, the conclusions we will reach are robust and do not
change qualitatively if the profile is distorted. 

To get the field equations, we vary the action with respect to the gauge 
field $A_\mu$ and the axion $\varphi$. The 
first variation yields the modified Maxwell equations, 
\be
\partial_\mu \Bigl(
F^{\mu\nu}
+ \frac{\zeta}{6 f_\phi}\,
\varphi\,\epsilon^{\mu\nu\lambda\sigma} F_{\lambda\sigma}
\Bigr) = 0 \, .
\label{Maxwellthick}
\ee
As noted, the Chern--Simons interaction 
contributes only when the axion field varies in spacetime. When $\varphi$ 
is constant, the additional term reduces to a total derivative and the 
equations locally reduce to those of vacuum electrodynamics.
The axion field equation is 
\be
\partial^2 \varphi = - \partial_\varphi V_{\tt I}(\varphi)
-\frac{\zeta}{4! f_\phi}\,
\epsilon^{\mu\nu\lambda\sigma} F_{\mu\nu}F_{\lambda\sigma} \, .
\label{axionfull}
\ee
In the regime of interest here we treat the axion as a fixed 
background and neglect the electromagnetic backreaction, 
which is consistent when the axion is supported by dynamics 
decoupled from the photon sector. Practically, this means that 
we linearize the field equations using the dynamical field fluctuations 
$F_{\mu\nu}$ and $\delta \varphi$ as small expansion parameters
around the background axion profile interpolating between two
adjacent axion vacua with vanishing electromagnetic background.
The background axion profile $\varphi_0$ is the solution of the homogeneous equation
$\partial^2 \varphi_0 = - \partial_\varphi V_{\tt I}(\varphi_0)$. For the
axion potential \eqref{sineG}, this equation can be solved exactly.
The domain wall profile is given by the sine-Gordon kink \cite{Vilenkin:2000jqa}
\be
2\pi\frac{\varphi_0(z)}{f_\phi}+\tilde\theta =
4\arctan\!\left(e^{m_\phi z/\sqrt{2}}\right) ,
\label{wallprofile}
\ee
where $m_\phi = 2\pi \mu^2 / f_\phi$ is the axion mass. 
This solution interpolates smoothly between distinct vacua over a finite 
distance set by the inverse axion mass,
\be
L \sim \frac{1}{m_\phi} \, ,
\ee
which defines the effective thickness of the wall. Clearly, for this to be valid,
we must have $m_\phi < {\cal M}$. Otherwise, we would have to integrate the
axion out and treat the wall as being ultrathin, 
describing the discharge of the dual top form flux ${\cal H}$ 
\cite{Kaloper:2026slg,Kaloper:2026ygk}.

The mode evolution is controlled by the gradient of the axion $\varphi$, which is
supported near the core of the wall,
\be
\frac{\zeta}{3 f_\phi}\,\partial_z \varphi_0(z)
= \frac{\zeta\, m_\phi}{3\sqrt{2}\,\pi \, \cosh \!\left(\frac{m_\phi z}{\sqrt{2}}\right)} \,  .
\label{gradphi}
\ee
The gradient vanishes at asymptotic infinity, where the background approaches constant $\theta$ 
regions. The axion gradient sets up the  
effective potential for electromagnetic waves. 
This potential provides a link to the sharp Chern--Simons interface 
considered previously. In the thin-wall limit, $L \to 0$, the smooth profile \eqref{gradphi}
reduces to a discontinuous jump, reproducing the $\delta$-function source for 
the Chern--Simons interaction, after we integrate the axion and turn on the top form 
dual ${\cal H}$. On the other hand, for finite $L$, the electromagnetic field 
propagates in a continuous chiral background determined by $\partial_z \varphi_0$.

\section{Disentangling Photon-Axion Fluctuations}

Let us now consider the field fluctuations in the domain wall background 
\eqref{wallprofile}. Both the background and the fluctuations are 
completely described by the full effective action \eqref{effaction}. It is important
to remember that this action is the result of considering all the
quantum corrections, perturbative or not, that involve the degrees of freedom
heavier than the photon and the axion, whose effects are included even after 
they are integrated out. This includes the heavier charged particles such
as electrons and (usual or dark) quarks, and nonperturbative contributions coming
from nontrivial vacuum topology below the chiral symmetry breaking in all
gauge sectors. 

Concretely, this means that all the relevant quantum corrections
that we need to concern ourselves with are already included in \eqref{effaction},
such as it is. Using it to account for those effects is therefore completely justified below
the cutoff of the action \eqref{effaction} roughly given by the mass of the lightest 
degree of freedom that was integrated out. Hence the action \eqref{effaction} is to
be treated as a purely classical but nonlinear theory of a background wall and 
electromagnetic and axionic fluctuations, that may or may 
not disturb the homogeneity of the wall \eqref{wallprofile}.

This theory contains three propagating degrees of freedom, 
the two photon helicities and an axion, which are massless and mass-gapped, respectively.
In particular the ``bulk" photon propagator $\langle 0 | T(A_\mu A_\nu) | 0 \rangle$ has a 
pole at $p^2 = 0$, which is protected by gauge redundancies enforced by the 
electromagnetic Ward identities. This persists in the presence of the wall, which breaks 
Poincar\'e symmetry spontaneously but does not affect gauge symmetries since the axion
bears no charge. Even if we were to include axion quantum corrections to the 
photon sector (that might arise due to the nontrivial wall background), 
because of the perturbative shift symmetry those would depend on $p^2$ and
would at most renormalize the residue of the photon pole, 
and perhaps dispersion relation coefficients, but would not shift it from 
zero or remove it. Thus it suffices for our  purposes
to consider only the classical phenomena that come from \eqref{effaction}. 

Turning to the details of symmetry breaking, the wall background \eqref{wallprofile} 
breaks the initial Poincar\'e symmetry $ISO(3,1)$ to
$ISO(2,1)$: the wall background is not invariant under $z$ translation, $z-t$ boost, and any rotations
involving the $z$-direction. Further we are interested in the scattering of electromagnetic waves on the
wall. Because $z-t$ boosts are broken, we can again use one to orient any incident wave along the
wall normal. Hence switching to this frame we can focus on 
gauge fields which define a subspace of the electromagnetic 
spectrum, with incoming wave vector parallel to the wall normal, $\vec k \parallel \hat z$. 

This subspace is
closed under evolution because the unbroken $ISO(2,1)$ includes translations and boosts along the wall, 
and so the components of the momentum along the wall are conserved. If they are zero in the incident state, 
they will stay zero all along. Thus if an incident wave does not depend on the wall coordinates
$\vec x_\parallel$ initially, scattering will not turn them on. To turn on those inhomogenous waves 
we must introduce them by changing initial conditions for incident waves far from the wall.

A similar argument shows that the axion field momenta parallel to the wall are also
conserved in scattering processes due to translational invariance along the wall.
If at higher order in perturbations axions are excited by incident gauge fields,
a homogeneous incident electromagnetic wave with vanishing parallel momentum
will not source axion modes with nonzero parallel momentum.
Moreover, if the incident electromagnetic frequencies satisfy $\omega < m_\phi$,
axion production is parametrically suppressed by $\omega/m_\phi < 1$. 
The axion mass serves as an effective UV cutoff 
of the localized mode sector. 
As we will see below, this is precisely the regime to which the analysis naturally
leads. Therefore we can consistently and reliably restrict the axion fluctuations to depend only
on $z,t$, limiting the analysis to the wall breathing modes. 

Combining this with gauge symmetries, which allows us to gauge-fix the relevant bulk modes 
to axial gauge that will be preserved by evolution, we can consistently use the 
ansatz $A^\mu = (0, \vec A_\perp, 0)$. As a result 
the field contents in the reduced theory invariant under $ISO(2,1)$ 
is still just a $2+1$-dimensional vector and a scalar, and we need not consider waves which are
inhomogenous along the wall. In the effective scattering
theory therefore there will be no mixing terms of the type $\propto \partial_z \varphi A_z$ and so on. 

Hence the evolution and scattering
of electromagnetic waves at the axion domain wall \eqref{wallprofile} remains effectively a 
$1+1$-dimensional problem even when walls are thick. The classicality of the action \eqref{effaction} 
excludes contributions from virtual states which have nonzero momenta transverse to the wall.

To extract the equations which govern our reduced problem, we 
return to Eqs. \eqref{Maxwellthick} and \eqref{axionfull}
and expand around the domain wall \eqref{wallprofile}, by turning on the gauge field $A_\mu$ 
and the axion fluctuation $\delta\varphi = \varphi - \varphi_0(z)$. Implementing the axial gauge
$A^\mu = (0, \vec A_\perp, 0)$ directly in the
full system of equations we rewrite \eqref{Maxwellthick} as
\be
\partial^2 A^\nu =  \frac{\zeta}{3 f_\phi} 
\partial_\mu \varphi \epsilon^{\nu\mu\lambda\sigma} \partial_\lambda A_\sigma \, ,
\label{fullmaxeqs}
\ee
and, with the symmetry decomposition outlined above,  after dropping the
subscript $\perp$,
\be
\partial^2
\begin{pmatrix}
A^x \\
A^y 
\end{pmatrix} + 
\frac{\zeta}{3f_\phi} \partial_z \varphi_0
\begin{pmatrix}
0 &1\\
-1& 0
\end{pmatrix}
\partial_t \begin{pmatrix}
A^x \\
A^y 
\end{pmatrix} = 
- 2 \frac{\zeta}{3f_\phi} \partial_{[z} \delta \varphi
\begin{pmatrix}
0 &1\\
-1& 0
\end{pmatrix}
\partial_{t]} \begin{pmatrix}
A^x \\
A^y 
\end{pmatrix} 
\, .
\label{mateq}
\ee
The terms on the left hand side are 
linear in the fluctuations, but depend on the 
nontrivial wall background, whereas the terms on the 
right hand side are genuine nonlinear interactions. 
On the right hand side, the indices enclosed by $[..., ...]$ are antisymmetrized, 
and the overall factor of 2 compensates the
usual 1/2 in the definition of antisymmetrization. 

Similarly, starting from the scalar field equation \eqref{axionfull}, 
we can split it into the linear and nonlinear pieces,
\be
\left(\partial^2 + V''(\varphi_0) \right)\delta\varphi =
-\frac{\zeta}{3 f_\phi} \Bigl(\partial_t A_x\,\partial_z A_y
- \partial_t A_y\,\partial_z A_x \Bigr) + {\cal W}(\delta \varphi, \varphi_0) .
\label{linAxion}
\ee
The latter terms schematically depicted by ${\cal W}$ are the scalar nonlinearities,
with terms at  least of order ${\cal O}(\delta \varphi^2)$. 

This system can be simplified by using circularly polarized vector fields  
\eqref{circcomp} as before. If we replace the vector field components
by linear combinations $A_\pm(z) = \frac{1}{\sqrt{2}} (A^x(z) \pm i A^y(z))$, 
the system splits into two gauge field subsystems, which 
only communicate with each other via the scalar exchange. The field equations
are
\ba
&& ~~\partial^2 A_\pm
\;\pm\; i\,\frac{\zeta}{3f_\phi}\,(\partial_z \varphi_0)\,\partial_t A_\pm
= \mp\, i\,\frac{\zeta}{3f_\phi}\, \Bigl(\partial_{z}\delta\varphi\,\partial_{t} A_\pm 
- \partial_{t}\delta\varphi\,\partial_{z} A_\pm \Bigr) \, , \nonumber \\
&&\left(\partial^2 + V_{\tt I}''(\varphi_0)
\right)\delta\varphi =  - \frac{i\zeta}{6 f_\phi}
\Bigl( \partial_t A_-\,\partial_z A_- - \partial_t A_+\,\partial_z A_+ \Bigr)
+ {\cal W}(\delta \varphi, \varphi_0) \, .
\label{nonlins}
\ea

From these equations we see that the relevant dimensionless parameters controlling the
dynamics are $\varphi_0/f_\phi$ and $\delta \varphi_0/f_\phi$. The former is set by 
the wall profile \eqref{wallprofile}, which means it is ${\cal O}(1)$ over the region 
set by the thickness of the wall, $\delta z \sim L \sim 1/m_\phi$. If the breathing mode
perturbation $\delta \varphi/f_\phi$ is nondestructive, $\delta \varphi/f_\phi <1$ 
so the wall does not fall apart by the perturbation, the right hand side of the first
of Eqs. \eqref{nonlins} is clearly subleading. 

Further, if a small electromagnetic pulse is sent in from infinity in a background 
which is initially a flat uniform wall without a $\delta \varphi$ excitation, as 
the pulse moves toward the wall, the second of Eqs. \eqref{nonlins} shows that it
excites the breathing mode, at order $\sim {\cal O}({A^2})$. This feeds back into the
propagation of the initial pulse, but as per the right hand side of the first of
Eqs. \eqref{nonlins}, only at order ${\cal O}(A^3)$. 

Finally, the right hand side terms in \eqref{nonlins} are both quadratic in derivatives, 
and so proportional to $\omega$ and $k$ of the initial state. Hence they only contribute the 
tiniest corrections in the IR to the poles of the propagation problem defined by \eqref{nonlins}. These 
terms can renormalize the phase and group velocities of the waves, but will not change the
power-law relation between $\omega$ and $k$ unless the perturbation is strong enough
to tear up the wall. 

This conclusion remains even if we allow the incident waves to come in at an angle 
and introduce inhomogeneities. To pursue their evolution, we'd have to go back to
more general equations \eqref{axionfull} and \eqref{fullmaxeqs}, but the same outcome 
would hold. As long as the wall perturbations are small, and the wall structure remains, the
theory's scattering properties are qualitatively the same at low momenta as the flat
wall limit, and the perturbations are but tiny perturbations of the smeared wall. After all,
that is precisely what one should expect to be protected by unbroken gauge symmetry.

This implies that to understand the spectrum of the theory with bulk electromagnetic 
fields and axions when a wall is added, we can resort to the restricted $1+1$-dimensional
problem with incident wave momenta which are parallel to $\hat z$ and the initial wall state is 
uniform. Then we can consistently ignore the 
nonlinear terms, looking only at the leading order electromagnetic 
wave scattering on the wall given by the left hand side terms in the first of the Eqs. 
\eqref{nonlins}. In other words, we solve for the leading order scattering matrix of
electromagnetic waves on a rigid axion domain wall of nonzero thickness. 
This problem is fully within the purview of quantum mechanics. 
There are perturbations outside of this regime, but those would 
destroy the wall, and so we ignore those. 

Said a bit more formally, to destabilize a localized electromagnetic 
mode, there should be a decay channel 
into propagating bulk waves. It would be indicated by an imaginary 
contribution to the frequency, signaling a finite lifetime. In our case no such 
channel will arise at leading order, since the dimension--$5$ operator does not
generate a mixing between the photon and $\delta\varphi$ that could allow the
bound state to mix with a continuum mode.

While the coupling $\delta\varphi\,\vec E\cdot\vec B$ provides a
mechanism whereby electromagnetic waves source axion perturbations, 
the ``emission rate" of axions $\delta \varphi$ is intrinsically suppressed by both 
the small amplitude of the wave and the derivative interactions, when wave momenta are
low. Moreover, the resulting
axion excitation must satisfy its own dispersion relation determined by the
effective potential $V_{\tt eff}$, which gaps it by the inverse wall thickness. This 
suppresses the kinematic phase space for
decay of the localized electromagnetic mode 
into propagating axions. If the axion mass approaches or exceeds
the effective cutoff, this channel shuts down completely, stiffening up the wall. 

\section{Scattering of Normally Incident Photons on the Wall}

As elaborated above, we now turn to the problem of determining the 
scattering matrix of photons normally incident on a flat wall, invariant under 
$ISO(2,1)$. The wall is defined by \eqref{wallprofile}, and the consistent
truncation of the field equations \eqref{nonlins} comprises of keeping just the
left hand side of the first of Eqs. \eqref{nonlins} and evaluating 
$\partial_z \varphi_0$ on \eqref{wallprofile}. Since the equation is also linear in the
gauge fields, we can consistently perform Fourier transform to the frequency domain,
$A_\pm = A_\pm e^{-i \omega t}$, and restrict the 
wave operator to $1+1$ dimensions, $\partial^2 = -(\partial_z^2 + \lambda)$.
This yields
\be
(\partial_z^2 + \lambda)\, A_\pm(z) = V_\pm(z)\, A_\pm(z)\, , \qquad \qquad 
V_\pm = \pm \,\frac{\beta_\pm}{\cosh\!\left(\gamma z\right)} \, . ~~~~
\label{Alins}
\ee
The parameters in the potential are
\be ~~~~
\gamma = \frac{m_\phi}{\sqrt{2}} \, , \qquad \qquad \beta_\pm 
= \frac{\zeta\, m_\phi \, \omega}{3\sqrt{2}\,\pi}\, .
\label{potparams}
\ee
We adopt the prescription for analytical continuation of the
eigenvalues of $\partial^2 + \lambda$ we discussed above,
and hold the power of frequency coming from 
Fourier transforming on the time coordinate fixed.
For our purposes it will suffice to explore the analytic structure of $(\partial_z^2 + \lambda)^{-1}$ 
and extract information about the poles -- i.e. about the existence of the normalizable
localized states -- and the dispersion relation
that this implies. 

Inverting \eqref{Alins} and rewriting it as an integral Lippmann-Schwinger equation amounts to 
inverting the operator $(\partial^2 + \lambda)$ using the outbound Green's function
$G(z - z') = -\frac{i}{2\sqrt{\lambda}}\, e^{i\sqrt{\lambda} |z - z'|}$. This gives
\be
A_\pm(z) = e^{i\sqrt{\lambda} z} 
- \frac{i}{2\sqrt{\lambda}} \int_{-\infty}^{\infty} dz'\,
e^{\,i\sqrt{\lambda} |z - z'|}\, V_\pm(z')\, A_\pm(z') \, ,
\label{LSvave}
\ee
where we normalized initial wave to unity, $A_\pm(-\infty) = 1$. 
Since the potential $V_{\pm}$ is not ultralocal anymore, evaluating this
equation cannot be done as simply as for the $\delta$-function potential. 
Nevertheless the integrability of the potential $\sim 1/\cosh(\gamma z)$
does allow us to find the leading order solution in controlled approximation. 
Since our main interest here is to prove that normalizable 
localized solutions \emph{exist}, this suffices.

Our strategy is to use analyticity of the bulk waves scattering on the 
smooth wall profile and approximate the \emph{exact} Lippmann-Schwinger equation
\eqref{LSvave} by the leading order term in the IR limit, with 
small frequencies $\sqrt{\lambda}$. This 
circumvents the perturbative Born approximation approach,
and enables us to calculate the
analogue of the ``gap" equation \eqref{A0sol} in the expansion 
$\sqrt{\lambda}/\gamma \sim \sqrt{\lambda}/m_\phi$, which will determine the
leading order term in the bulk wave scattering matrix in $\sqrt{\lambda}/m_\phi$,
that controls the leading order localized wave dispersion relation and confirms
the existence of the localized wave. 
The higher order corrections can in principle shift the coefficients in the 
dispersion relation. However, just resuming the perturbative series of higher order terms in our
classical perturbation theory as described in the previous section will
not remove these modes altogether. 

Turning our attention back to \eqref{LSvave}, we focus on the limiting form when
$z \rightarrow \infty$, which implies $z > z'$ inside the integral. This follows since 
the potential $V_\pm = \pm \,\frac{\beta_\pm}{\cosh\!\left(\gamma z\right)}$ is mostly supported
in the interval $-\frac{1}{\gamma} < z' < \frac{1}{\gamma}$, which can be made arbitrarily narrow by
taking $\gamma \sim m_\phi$ large. In this case the integral in \eqref{LSvave} is 
approximated by
\be
\int_{-\infty}^{\infty} dz'\,
e^{\,i\sqrt{\lambda} |z - z'|}\, V_\pm(z')\, A_\pm(z') \simeq 
e^{\,i\sqrt{\lambda} z} \int_{-1/\gamma}^{1/\gamma} dz'\,
e^{-\,i\sqrt{\lambda}z'}\, V_\pm(z')\, A_\pm(z')\, .
\label{LSint}
\ee
The relative error of this approximation is roughly given by the ratio of $\sqrt{\lambda}/\gamma$
essentially due to the exponential suppression 
of $V_\pm$ at $|z| > 1/\gamma$, and the destructive interference
of the wave outside of the wall as $z \rightarrow \infty$. The second approximation is 
similar; since we are interested in the
IR regime of \eqref{LSvave}, we can approximate the exact wave profile $A_\pm(z')$ inside the
wall by it's central value times an overall phase controlled by $\sqrt{\lambda}$. In other words, we simply
replace a scattering wave packet with frequencies centered around $\sqrt{\lambda}$ by a monochromatic
wave, ignoring the spreading, $A_\pm(z') \simeq A_\pm(0) e^{i\sqrt{\lambda}z'}$, where the sign
of the phase reflects our boundary condition that the wave is outgoing from left to right. This 
introduces additional approximation errors, but still organized as the 
expansion in  $\sqrt{\lambda}/\gamma$. Thus, we find 
\ba
&& e^{\,i\sqrt{\lambda} z} \int_{-1/\gamma}^{1/\gamma} dz'\,
e^{-\,i\sqrt{\lambda}z'}\, V_\pm(z')\, A_\pm(z') \simeq e^{\,i\sqrt{\lambda} z} \int_{-1/\gamma}^{1/\gamma} dz'\,
e^{-\,i\sqrt{\lambda}z'}\, V_\pm(z')\, A_\pm(0) e^{i\sqrt{\lambda}z'} ~~~ \nonumber \\
&& \qquad \qquad \qquad \qquad \qquad \qquad \qquad ~~~
\simeq e^{\,i\sqrt{\lambda} z} \int_{-\infty}^{\infty} dz'\, V_\pm(z')\, A_\pm(0) \, , ~~~
\label{LSint2}
\ea
where in the last step we extended the limits of 
integration to the whole real line, which is consistent since
the error remains to be ${\cal O}(\sqrt{\lambda}/\gamma)$. Plugging this into \eqref{LSvave} yields
\be
A_\pm(z) = e^{i\sqrt{\lambda} z} \Bigl\{1 
- \frac{i}{2\sqrt{\lambda}} \Bigl(\int_{-\infty}^{\infty} dz'\, V_\pm(z')\Bigr) \,
A_\pm(0) \Bigr\} +  {\cal O}\bigl(\frac{\sqrt{\lambda}}{\gamma} \,\bigr)  \, .
\label{LSvaveTr}
\ee

Because the potential $V_\pm$ is integrable, 
\be
\int_{-\infty}^{\infty} dz'\, V_\pm(z')\, =
\frac{\pi \beta_\pm}{\gamma} \, .
\label{potint}
\ee
Substituting this into Eq. \eqref{LSvaveTr}, evaluating it at $z=0$ and 
solving for $A_\pm(0)$ yields the gap equation of the same formal structure as
\eqref{A0sol}, 
\be
A_\pm(0)=\frac{1}{1 + \frac{i\pi \beta_\pm}{2\gamma\sqrt{\lambda}}} 
+ {\cal O}\bigl(\frac{\sqrt{\lambda}}{\gamma} \,\bigr) \, .
\label{A0solthick}
\ee
Using $\gamma = m_\phi/\sqrt{2}$ and 
$\beta_\pm = \frac{\zeta\, m_\phi \, \omega}{3\sqrt{2}\,\pi}$, we recall that the distance between
adjacent axion vacua is $\Delta \varphi/f_\phi = 1$ (see Eq.\ \eqref{vacuadist}). 
Matching the normalization of the axion electrodynamics coupling in Eq.\ \eqref{action} 
to the Chern--Simons interface description employed in the thin-wall limit, and taking into account 
the normalization of the Green's function in the scattering problem, we identify
\be
\Delta \theta = \frac{\zeta}{3} \, ,
\label{deltheta}
\ee
which ensures consistency with the phase shift 
extracted from the thin-wall scattering solution.
Substituting in Eq. \eqref{A0solthick}, 
we see that the poles of the gap equation \eqref{A0solthick}
are located precisely at
\be
1 \pm \frac{i \Delta \theta \omega}{2\sqrt{\lambda}} 
+  {\cal O}\bigl(\frac{\sqrt{\lambda}}{\gamma} \,\bigr) =  0 \, .
\label{thickpoles}
\ee
Analytically continuing to $\sqrt{\lambda} = i \kappa$, just like above, and noting that
the incident wave prescription then implies $\exp(i \sqrt{\lambda}z) \rightarrow \exp(- \kappa z)$,
this means that the normalizable localized mode corresponds to 
\be
\kappa = \frac{\Delta\theta }{2}  \omega 
+  {\cal O}\bigl(\frac{\sqrt{\lambda}}{\gamma} \,\bigr) {\rm ~~corrections}\, . 
\label{kappapoleth}
\ee
This is precisely the left-handed, thin wall chiral surface mode 
corrected by the ${\cal O}({\sqrt{\lambda}}/{\gamma})$
terms which arise in the thick wall backgrounds, and disappear when the axion 
decouples as $m_\phi \sim \gamma \rightarrow \infty$. This precisely confirms our claim that the 
thick walls {\it also} support the localized normalizable electromagnetic waves.

Note that our formula \eqref{LSvaveTr} is an approximate solution for the full transmitted 
wave function because it is obtained from taking the exact Lippmann-Schwinger 
solution \eqref{LSvave} in the regime $z \rightarrow \infty$, where one calculates the 
fraction of the incident wave that passed through the wall. This is echoed in the factorized
wave phase term in \eqref{A0solthick}. If we then restrict \eqref{LSvave} to the first term in the
Born approximation,
\be
A_\pm(z) = e^{i\sqrt{\lambda} z} 
- \frac{i}{2\sqrt{\lambda}} \int_{-\infty}^{\infty} dz'\,
e^{\,i\sqrt{\lambda} |z - z'|}\, V_\pm(z')\, e^{i\sqrt{\lambda} z'} \, ,
\label{LSvaveborn}
\ee
the approximate solution reduces to
\be
A_\pm(z) = e^{i\sqrt{\lambda} z} \Bigl\{1 
- \frac{i}{2\sqrt{\lambda}} \Bigl(\int_{-\infty}^{\infty} dz'\, V_\pm(z')\Bigr) \,
\Bigr\} +  {\cal O}\bigl(\frac{\sqrt{\lambda}}{\gamma} \,\bigr) \, .
\label{LSvaveTrb}
\ee
The term multiplying the phase factor on the right hand side is precisely the leading order 
expression for the transmission amplitude, 
\be 
T_\pm(\sqrt{\lambda}) = 1 \mp 
\frac{i \Delta \theta \omega}{2\sqrt{\lambda}}  +  {\cal O}\bigl(\frac{\sqrt{\lambda}}{\gamma} \,\bigr)   \, ,
\label{transamp} 
\ee
after replacing $\gamma$ and $\beta_\pm$ using \eqref{potparams}, and again using
$\Delta \theta \equiv \frac{\zeta}{3}$. For bulk modes for which
$\sqrt{\lambda} = \omega$, substituting $\gamma = m_\phi/\sqrt{2}$, this yields 
\be
T_\pm(\sqrt{\lambda}) = 1 \mp 
\frac{i \Delta \theta}{2}  + {\cal O}\bigl(\frac{\omega}{m_\phi} \bigr) 
=  1 \mp \frac{i \zeta}{6}  + {\cal O}\bigl(\frac{\omega}{m_\phi} \bigr) \, .
\label{pancha}
\ee
The leading order term is precisely the Pancharatnam phase measuring the 
rotation of the polarization plane
of light traversing the domain wall in the regime when $\zeta/6$ 
is small \cite{Pancharatnam:1956url,Pancharatnam:1956url2,Born:1999ory}, 
which we have calculated exactly in the thin wall limit in 
\cite{Kaloper:2026ygk}. Here we obtained this result in an ``en passant" manner, 
because it is inextricably 
linked to the existence of the normalizable chiral electromagnetic wave 
localized to the wall. 

We need to stress here that our analysis is \emph{complementary} to the 
past work, because it explores the regime which is \emph{not} 
covered by the adiabatic approximation employed in previous 
analyses of axion-induced birefringence and electromagnetic wave propagation
in axion field backgrounds, as exemplified in \cite{Harari:1992ea,Huang:1985tt}
and subsequent work. In those treatments, 
a bulk electromagnetic wave is assumed to propagate 
through a slowly varying axion background with frequency much larger than the axion 
mass, $\omega \gg m_\phi$ \cite{Harari:1992ea,Huang:1985tt,Raffelt:1987im}. 
In that regime, one treats the axion 
profile as approximately constant over a wavelength. This leads to a 
local description of polarization rotation as a cumulative effect along 
the trajectory. Although this approach captures the leading birefringence 
phenomenon, it does not probe the full spectral structure of the wave 
operator in the presence of spatial gradients. In particular, the adiabatic regime is
not well equipped to uncover the localized modes on the wall. Our result shows
that those modes are cleanly extracted in the limit when $\omega < m_\phi$, which is 
precisely complementary with the conventional adiabatic perturbation theory. 

The localized wave dynamics remains under perturbative control precisely 
where the adiabatic approximation fails. The axion mass is an effective UV cutoff 
of the localized mode sector. It defines the inverse thickness of the wall and
sets the highest frequency at which electromagnetic waves see the wall as a
coherent background. For $\omega > m_\phi$, the wall is resolved at the scale
of its internal structure, and the localized mode description begins to melt into
continuum scattering states. In the Lippmann--Schwinger framework, 
the pole in the analytically continued resolvent will be strongly dressed by the
corrections which accumulate as the condition
$\omega < m_\phi$ is violated. This pole should 
be understood as a low-energy bound-state pole of the effective theory. 
It is sharp throughout the controlled regime $\omega < m_\phi$. Outside of this
regime, the state which this pole corresponds to begins to dissolve 
into the continuum as this condition is relaxed.

Thus the chiral surface photon and the associated phase shift are 
intrinsic features of the full non-perturbative spectral 
problem. Our results complete the connection 
between axion-induced birefringence and interface dynamics 
by extending it beyond the conventional adiabatic regime. This regime is  
fully under control thanks to the analyticity of the scattering processes
and the symmetries of the system. Further, the localized mode
survives in the limit when the axion mass is pushed above the UV cutoff of the
photon-axion theory \eqref{effaction}, $m_\phi > {\cal M}$, and the effective
picture of the wall reduces to a thin $\delta$-function limit of \eqref{actioncs}. 

From the effective field theory vantage point, the presence of the localized mode
is tied directly to the regime in which the domain wall behaves as a coherent
background. The condition $\omega < m_\phi$ ensures that
electromagnetic waves do not resolve the internal structure of the wall, but see it as 
a ``whole" rather than ``a sum of the parts", and propagate in an effective medium 
supporting a bound state. The loss of
the wall's ``rigidity" at scales $\omega > m_\phi$ will eventually lead to the 
dissolution of the bound mode into the continuum spectrum.

There are immediate implications of our results beyond establishing the control
of the non-adiabatic regime of the axion domain wall theory. First and foremost,
axion domain walls may capture electromagnetic energy and store it in the form
of localized electromagnetic waves propagating along them. Such energy could get off the
wall if the wall is large but finite, or, particularly in cosmology, if the wall breaks down. These types 
of phenomena could lead to completely new axion domain wall signatures, 
which have not been anticipated to date. These phenomena could also lead to novel types of
waveguides that might be realizable in condensed matter systems. Moreover, the
analysis here also shows that the thin wall chiral electromagnetic surface waves found
in \cite{Kaloper:2026auc} are not just some special features of ultrathin, $\delta$-function 
domain wall structures, but are robust consequences of regions which support 
the Maxwell Chern--Simons terms. We further note that while we have 
worked here with the simple description of the background axion domain 
wall \eqref{wallprofile}, our central result summarized in Eq. \eqref{LSvaveTr} only 
relies on the feature that wall has finite thickness. One would therefore expect 
qualitatively the same type of normalizable localized electromagnetic
walls for other type of axion potentials. 

\section{Summary}

We have shown here that Maxwell theory coupled to axion domain walls
admits a robust and previously unrecognized electromagnetic wave localized
on the walls. In the ultrathin limit, where the axion profile reduces to a
Chern--Simons interface, we recover the single chiral surface
photon, characterized by exponential localization in the transverse direction
and linear, gapless dispersion along the wall, 
which we found recently in \cite{Kaloper:2026auc}. This mode arises from the
helicity-dependent interaction induced by the axion gradient, which acts as an
attractive $\delta$-function potential for one polarization and repulsive for
the other. The helicity which experiences attractive potential 
supports a normalizable electromagnetic bound state.

To show this we used the Lippmann--Schwinger approach to scattering,
demonstrating that the localized mode can be directly inferred from the analytic
structure of the bulk photon scattering amplitude. The bound 
surface photon is identified as a pole of the analytically continued resolvent,
located on the evanescent sheet of the complex momentum plane. This shows
that the bound state is not just an additional degree of freedom introduced by hand,
but is already contained in the propagation of bulk electromagnetic modes in the
presence of the interface. The same analytical structure simultaneously supports both 
delocalized mode birefringence and the chiral mode localization, revealing them as
two manifestations of a single dynamical mechanism.

A central result of our work is that the surface mode is not an
artifact of the $\delta$-function approximation. By extending the analysis to
axion domain walls of finite thickness we have shown that the pole in the scattering 
amplitude persists for generic resolved walls in the effective field theory below the cutoff. 
The localized electromagnetic wave survives continuous deformations of the 
background, that can be parameterized by axion 
field fluctuations, with corrections organized in powers
of $\sqrt{\lambda}/m_\phi$. Thus the $\delta$-function description 
emerges as the controlled thin-wall limit of a
much wider phenomenon. This makes contact with analogous interface
modes familiar from optical and condensed 
matter systems \cite{Born:1999ory,Economou:1969zz}. 

To summarize, our derivation is completely complementary to the adiabatic
approximation used in previous analyses of axion-induced
birefringence starting with 
\cite{Harari:1992ea,Huang:1985tt}. In those treatments, one assumes
that electromagnetic waves propagate through a slowly varying axion background
with frequency $\omega \gg m_\phi$. In our approach the 
electromagnetic propagation is treated by studying the wave equation 
and its resolvent, in the complementary regime $\sqrt{\lambda} 
\sim \omega  < m_\phi$. The corrections
to the thin-wall limit scale as $\omega/m_\phi$, and so they remain small
precisely where the adiabatic approximation breaks down.
Thus our results complete the correspondence between
axion-induced birefringence and interface dynamics by extending it beyond 
the adiabatic regime. In addition, we find the 
localized electromagnetic mode which has been hidden
to the adiabatic explorations all along.

Our results identify a new
class of topological electromagnetic phenomena associated with axion
backgrounds. It would be interesting to seek explicit solutions for the
localized modes in smooth wall profiles, to study their interactions, and to
investigate possible observational signatures of such surface photons in
astrophysical/cosmological environments and also in condensed matter 
realizations of axion-like defects.


\end{document}